\documentclass[preprint,12pt]{elsarticle}

\usepackage[utf8]{inputenc}
\usepackage[T1]{fontenc}

\usepackage{amsmath}
\usepackage{amsfonts}
\usepackage{amssymb}
\usepackage{amsthm}
\usepackage{graphicx}
\usepackage{enumerate}
\usepackage{bm}
\usepackage{booktabs}
\usepackage{multirow}
\usepackage{mathrsfs}

\graphicspath{{Figures/}}

\begin{document}

\title{Nonlinear waves in solids with slow dynamics: an internal-variable model}

\author{
H.~Berjamin$^{1}$, N.~Favrie$^{2}$, B.~Lombard$^{1}$ and G.~Chiavassa$^{3}$}

\address{$^{1}$Aix Marseille Univ, CNRS, Centrale Marseille, LMA, Marseille, France\\
$^{2}$Aix-Marseille Univ, UMR CNRS 7343, IUSTI, Polytech Marseille, 13453 Marseille Cedex 13, France\\
$^{3}$Centrale Marseille, CNRS, Aix-Marseille Univ, M2P2 UMR 7340, 13451 Marseille Cedex 20, France
}




\begin{abstract}
	In heterogeneous solids such as rocks and concrete, the speed of sound diminishes with the strain amplitude of a dynamic loading (softening). This decrease known as ``slow dynamics'' occurs at time scales larger than the period of the forcing. Also, hysteresis is observed in the steady-state response. The phenomenological model by Vakhnenko et al. is based on a variable that describes the softening of the material [{\itshape Phys.~Rev.~E} {\bfseries 70}-1, 2004]. However, this model is 1D and it is not thermodynamically admissible. In the present article, a  3D model is derived in the framework of the finite strain theory. An internal variable that describes the softening of the material is introduced, as well as an expression of the specific internal energy. A mechanical constitutive law is deduced from the Clausius-Duhem inequality. Moreover, a family of evolution equations for the internal variable is proposed. Here, an evolution equation with one relaxation time is chosen. By construction, this new model of continuum is thermodynamically admissible and dissipative (inelastic). In the case of small uniaxial deformations, it is shown analytically that the model reproduces qualitatively the main features of real experiments.
\end{abstract}


	


\maketitle



\section{Introduction}

Rocks and concrete are known to have a strong nonlinear behaviour. Quasistatic compression or traction tests show a nonlinear stress-strain relationship. A hysteresis loop is observed when the loading is increased and decreased. This phenomenon is interpreted as a memory effect~\cite{guyer99}. The longitudinal vibrations of a rod of material also show highly nonlinear features. Indeed, a frequency shift of resonance peaks is observed when the amplitude of the vibration is increased. The frequency shift reveals a global softening of the material with the strain amplitude~\cite{guyer99,tencate11}.

\begin{figure}
	\centering
	\includegraphics[width=0.97\textwidth]{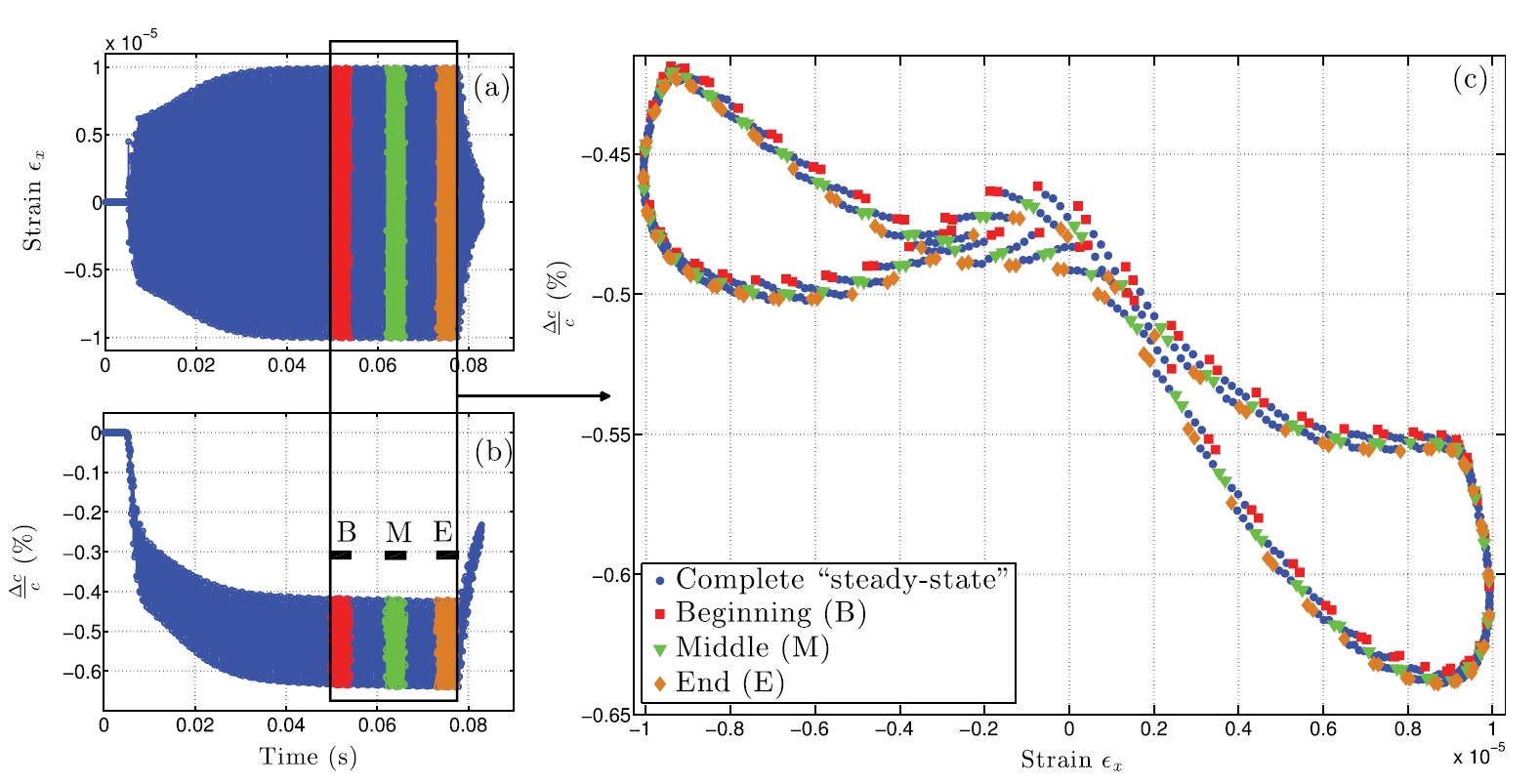}
	
	\caption{Dynamic acoustoelasticity measurement. (a) Evolution of the axial strain $\varepsilon$ at the location of the probe over time. (b) Relative variation $\Delta c/c$ of the sound speed with respect to its initial value over time. (c) Hysteresis loop: $\Delta c/c$ versus $\varepsilon$ in steady state. Reproduced from \cite{riviere13}.
	\label{fig:RiviereSoftening}}
\end{figure}

In dynamic acoustoelastic testing (DAET), the speed of sound is measured locally over time, when longitudinal vibrations are simultaneously applied to the whole sample. As illustrated on figure~\ref{fig:RiviereSoftening}-(b), a decrease with time of the measured sound speed is observed. This \emph{softening} occurs over a time scale larger than the period of the dynamic loading, which highlights the phenomenon of \emph{slow dynamics}. Moreover, the evolution of this speed with respect to the strain presents an hysteresis curve in steady state (figure~\ref{fig:RiviereSoftening}-(c)). When the excitation is stopped ($t \approx 0.08$~s in figure~\ref{fig:RiviereSoftening}-(b)), the sound speed increases, and recovers gradually its initial value (\emph{recovery}). All these phenomena are accentuated when the strain amplitude is increased~\cite{riviere13}.

Several dynamic models which reproduce these features can be found in the literature~\cite{meurer02,li15}. One approach consists in incorporating a dependency on the strain rate in the stress-strain relationship~\cite{nazarov03}. Another approach is related to the Preisach-Mayergoyz model, which is based on a discrete representation of hysteresis~\cite{mccall94a,meurer02,abeele04}. The \emph{soft-ratchet model} of Vakhnenko et al. results from a different approach~\cite{vakhnenko04,lombardWM15}. A new variable $g$ is introduced so as to describe the softening. Interpreted as a concentration of activated defects, this variable modifies the apparent elastic modulus. Also, an evolution equation for $g$ is provided. In this equation, a relaxation time is incorporated to describe the slow dynamics. Nevertheless, the soft-ratchet model was developed in one space dimension, and does not generalize straightforwardly to higher space dimensions. Moreover, thermodynamical issues are not considered in the construction of this model.

In the present article, a new phenomenological model is proposed in the context of the finite strain theory. Similarly to the soft-ratchet model, a scalar internal variable $g$ is introduced to describe the softening. Our model, derived from beginning in the framework of continuum mechanics with internal variables~\cite{maugin94a,maugin15}, satisfies by construction the principles of thermodynamics.
As shown later in the document, a particular choice of the internal energy yields a separable constitutive law
$$
\bm{\sigma} = \left(1-g\right) \bar{\bm{\sigma}}(\bm{\chi}) \, ,
$$
where $\bm{\sigma}$ is the Cauchy stress and $\bm{\chi}$ is a strain tensor. Such a constitutive law resembles classical models of irreversible damage. Furthermore, an evolution equation for the internal variable of the form
$$
\dot{g} = \mathscr{S}(\bm{\chi},g)
$$
is obtained, where $\dot{g}$ denotes the material derivative of $g$. Here, both $\dot{g} \geqslant 0$ and $\dot{g}\leqslant 0$ are possible. If $\dot{g} \geqslant 0$, the sound speed proportional to $\sqrt{1-g}$ decreases (softening). Inversely, $\dot{g} \leqslant 0$ increases the sound speed (recovery). In the choice of the evolution equation $\mathscr{S}(\bm{\chi},g)$, particular care is taken to ensure that the Clausius-Duhem inequality is satisfied whatever the sign of $\dot{g}$. This is a major difference with damage modelling, where the internal variable $g$ describes an irreversible process, so that only $\dot{g} \geqslant 0$ is possible~\cite{holzapfel00}.

The article is organized as follows. In section~\ref{sec:Model}, the model is constructed, leading to the constitutive law and the evolution equation. Several examples of finite-strain models are provided for illustration purposes. In particular, the cases of infinitesimal strain and uniaxial strain are addressed. In section~\ref{sec:AnalysisModel}, the equations are solved analytically in a particular configuration. The three expected phenomena{\,---\,}softening, slow dynamics and hysteresis{\,---\,}are reproduced by the model.
In appendix~\ref{sec:Extensions}, a link is made between the new model and quasistatic models of filled rubber. Moreover, a formal analogy with a system of wet sticking fibers is proposed. Lastly, in appendix~\ref{sec:Vakhnenko}, we demonstrate that the soft-ratchet model originally proposed by Vakhnenko et al. is not thermodynamically relevant.


\section{Construction of the model}\label{sec:Model}

\subsection{Basic equations}

Let us consider an homogeneous continuum on which no external volume force is applied, and no heat transfer occurs. Furthermore, self-gravitation is neglected. A particle initially located at some position $\bm{x}_0$ of the reference configuration moves to a position $\bm{x}_t$ of the current configuration. The deformation gradient is a second-order tensor defined by (see e.g. \cite{ogdenElast84,holzapfel00,norris98})
\begin{equation}
	\bm{F} = \mathbf{grad}\, \bm{x}_t = \bm{G} +  \mathbf{grad}\, \bm{u}\, ,
	\label{F}
\end{equation}
where $\bm{u} = \bm{x}_t - \bm{x}_0$ denotes the displacement field and $\mathbf{grad}$ is the gradient with respect to the material coordinates $\bm{x}_0$ (Lagrangian gradient). In the reference configuration, the deformation gradient (\ref{F}) is equal to the metric tensor $\bm{G}$. If the Euclidean space is described by an orthonormal basis and a Cartesian coordinate system, the matrix of the coordinates of $\bm G$ is the identity matrix.

The choice of a representation of motion{\,---\,}Eulerian or Lagrangian{\,---\,}does not affect the expressions of the constitutive laws and the evolution equations. However, it affects the expression of the material derivative and the equations of motion. Here, the Lagrangian representation of motion is used. Hence, the material derivative $\dot{\bm \psi}$ of any field $\bm{\psi}(\bm{x}_0,t)$ is
\begin{equation}
	\dot{\bm{\psi}} = \frac{\partial \bm{\psi}}{\partial t}\, .
	\label{MaterialDeriv}
\end{equation}
In particular, the material derivative of the deformation gradient satisfies
\begin{equation}
	\dot{\bm{F}} = \mathbf{grad}\,\bm{v} \, ,
	\label{FDotGen}
\end{equation}
where $\bm{v}(\bm{x}_0,t)$ is the velocity field. The conservation of mass implies
\begin{equation}
	\frac{\rho_0}{\rho} = \det(\bm{F}) \, ,
	\label{ConsMassF}
\end{equation}
where $\rho$ denotes the mass density in the deformed configuration, and $\rho_0$ denotes the mass density in the reference configuration. The motion is also driven by the conservation of momentum
\begin{equation}
	\rho_0\, \dot{\bm{v}} = \mathbf{div} \left(\det(\bm{F})\, \bm{\sigma}\cdot\bm{F}^{-\top}\right) ,
	\label{ConsMom}
\end{equation}
where $\mathbf{div}$ denotes the divergence with respect to the material coordinates. The expression of the Cauchy stress tensor $\bm \sigma$ will be specified later on.

As usual in acoustics, the thermodynamic process is assumed to be adiabatic. The first principle of thermodynamics introduces the specific internal energy $e$. The conservation of total energy writes:
\begin{equation}
	\rho\,\dot{e} = \bm{\sigma} : \bm{D}\, ,
	\label{Thermo1}
\end{equation}
where $\bm{D} = \frac{1}{2} (\dot{\bm{F}}\cdot \bm{F}^{-1} + \bm{F}^{-\top}\!\cdot\dot{\bm{F}}^\top )$ is the strain-rate tensor. The second principle of thermodynamics reads
\begin{equation}
	\rho\,\dot{s} \geqslant 0\, ,
	\label{Thermo2}
\end{equation}
where $s$ is the specific entropy.

\subsection{The model}

\paragraph{Preliminaries.}
We choose the following variables of state: the specific entropy $s$, the strain tensor $\bm\chi$, and an additional scalar variable $g$, which is introduced to represent the softening/recovery of the material. Consequently, the Gibbs identity reads
\begin{equation}
\dot{e} = T\,\dot{s}
+ \left.\frac{\partial e}{\partial \bm{\chi}}\right|_{s, g} :\dot{\bm \chi}
+ \left.\frac{\partial e}{\partial g}\right|_{s, \bm{\chi}} \dot{g}\, ,
\label{Gibbs}
\end{equation}
where $T = \left.{\partial e}/{\partial s}\right|_{\bm{\chi}, g} > 0$ is the absolute temperature. Multiplying (\ref{Gibbs}) by $\rho$, the local equations of thermodynamics (\ref{Thermo1})-(\ref{Thermo2}) yield the Clausius-Duhem inequality
\begin{equation}
\mathscr{D} = \bm{\sigma} : \bm{D}
- \rho\!\left.\frac{\partial e}{\partial \bm{\chi}}\right|_{s, g} : \dot{\bm \chi}
- \rho\!\left.\frac{\partial e}{\partial g}\right|_{s, \bm{\chi}} \dot{g}
\geqslant 0\, ,
\label{Thermo2Diss}
\end{equation}
for all state $\lbrace s, \bm{\chi}, g\rbrace$ and all evolution $\lbrace\dot s, \dot{\bm \chi}, \dot g\rbrace$. The left-hand term in (\ref{Thermo2Diss}) is the dissipation $\mathscr{D}$ per unit volume of material (W.m\textsuperscript{-3}).

The main ingredient of the model is an expression of the internal energy per unit volume of the form
\begin{equation}
	\rho_0\, e = \phi_1(g)\, W(\bm{\chi}) + \phi_2(g)\, ,
	\label{ConstitutiveE}
\end{equation}
where $W$ is the strain energy density function, expressed in terms of the strain tensor $\bm{\chi}$. The function $\phi_1$ has dimensionless values, and $\phi_2$ is a storage energy. If $\phi_1(g) = 1$ and $\phi_2(g) = 0$ for all $g$, then the classical case of hyperelasticity is recovered, where $\rho_0\, e = W(\bm{\chi})$. The expression of the internal energy (\ref{ConstitutiveE}) is analogous to the Ogden-Roxburgh model of filled rubber~\cite{ogden99}. It is also formally analogous to a model of wet sticking fibers~\cite{py07a}. These similarities are detailed in appendix~\ref{sec:Extensions}.

With the assumption (\ref{ConstitutiveE}), the following substitutions are made in the inequality (\ref{Thermo2Diss}):
\begin{equation}
{\addtolength{\jot}{0.4em}
	\begin{aligned}
	&\rho\!\left.\frac{\partial e}{\partial \bm{\chi}}\right|_{s, g} : \dot{\bm \chi} = \phi_1\frac{\rho}{\rho_0}\,\frac{\partial W}{\partial \bm{\chi}} : \dot{\bm \chi} \, ,\\
	&\rho\!\left.\frac{\partial e}{\partial g}\right|_{s, \bm{\chi}}\dot{g} = \frac{\rho}{\rho_0}\left(\phi'_1\, W + \phi'_2\right)\dot{g}\, ,
	\end{aligned}}
\label{Thermo2DissSubs}
\end{equation}
where $\phi'_1$ and $\phi'_2$ denote the derivatives of $\phi_1$ and $\phi_2$, respectively. The final constitutive laws are obtained for a given choice of strain tensor $\bm\chi$. In the next paragraph, the right Cauchy-Green tensor $\bm{C} = \bm{F}^\top\! \cdot \bm{F}$ is used. For many other strain tensors, the constitutive laws can be deduced from $\bm C$, and similar derivations can be done.

\paragraph{Constitutive laws.}

We choose the right Cauchy-Green tensor $\bm{\chi} = \bm{C} = \bm{F}^\top\! \cdot \bm{F}$. The material derivative of the strain tensor is $\dot{\bm{C}} = 2\bm{F}^\top\!\cdot\bm{D}\cdot\bm{F}$. For any second-order tensors $\bm T$, $\bm F$ and $\bm D$, we recall that
\begin{equation}
	{\addtolength{\jot}{0.4em}
	\begin{aligned}
		\bm{T} : (\bm{F}^\top \cdot \bm{D} \cdot \bm{F}) &= (\bm{D} \cdot \bm{F}) : (\bm{F} \cdot \bm{T}) \\
		&= \mathrm{tr} (\bm{D} \cdot \bm{F} \cdot (\bm{F} \cdot \bm{T})^\top)\\
		&= \mathrm{tr} (\bm{D} \cdot (\bm{F} \cdot \bm{T} \cdot \bm{F}^\top)^\top)\\
		&= (\bm{F} \cdot \bm{T} \cdot \bm{F}^\top) : \bm{D} \, .
	\end{aligned}}
	\label{ManipTensor}
\end{equation}
Therefore, the Clausius-Duhem inequality (\ref{Thermo2Diss}) with the substitutions (\ref{Thermo2DissSubs}) reduces to
\begin{equation}
	\mathscr{D} = \underbrace{\left(\bm{\sigma}-\phi_1\,\bar{\bm \sigma}\right) : \bm{D}}_{\mathscr{D}_\mathit{el}} \underbrace{- \rho/\rho_0\left(\phi_1'\,W + \phi_2'\right) \dot{g}}_{\mathscr{D}_\mathit{inel}} \geqslant 0\, ,
	\label{Thermo2DissF}
\end{equation}
where the hyperelastic stress
\begin{equation}
	\bar{\bm \sigma} = \frac{\rho}{\rho_0}\, \bm{F}\cdot 2\frac{\partial W}{\partial \bm{C}} \cdot \bm{F}^\top
	= \frac{1}{\det(\bm{F})}\, \bm{F}\cdot 2\frac{\partial W}{\partial \bm{C}} \cdot \bm{F}^\top
	\label{StressF}
\end{equation}
depends on $\bm F$.

The stress $\bm \sigma$ is a state function: it does not dependent on $\bm D$, which is not a variable of state. Thus, the term $\mathscr{D}_\mathit{el}$ in the dissipation (\ref{Thermo2DissF}) is a scalar product between $\bm D$ and a tensor which does not depend on $\bm{D}$. Moreover, the term $\mathscr{D}_\mathit{inel}$ does not depend on $\bm D$. Therefore, the Clausius-Duhem inequality (\ref{Thermo2DissF}) for all $\bm D$ yields the constitutive law 
\begin{equation}
	\bm{\sigma} = \phi_1(g)\, \bar{\bm\sigma}(\bm{\chi}) \, ,
	\label{ConstitutiveEq}
\end{equation}
where the hyperelastic stress $\bar{\bm\sigma}$ is defined in (\ref{StressF}).

Now, the Clausius-Duhem inequality (\ref{Thermo2DissF}) reduces to $\mathscr{D}_\mathit{inel} \geqslant 0$, for all state and all $\dot g$. Therefore, $\phi_1'\,W + \phi_2'$ is either dependent on $\dot g$ or equal to zero. We choose the simplest nontrivial dependence:
\begin{equation}
	\phi_1'\,W + \phi_2' = -\tau_1\, \dot{g} \, ,
	\label{RelG}
\end{equation}
where $\tau_1 = \tau\times 1$~J/m\textsuperscript{3} and $\tau \geqslant 0$ is a relaxation time. The parameter $\tau$ may be variable, e.g. dependent on the sign of $\dot{g}$, temperature, or any desired parameter. If $\tau\neq  0$, equation (\ref{RelG}) gives the evolution equation
\begin{equation}
	\dot{g} = -\frac{1}{\tau_1}\left(\phi'_1(g)\, W(\bm{\chi}) + \phi_2'(g)\right) .
	\label{EvolG}
\end{equation}
Otherwise ($\tau = 0$), the internal variable $g$ satisfies $\phi_1'\,W + \phi_2' = 0$, i.e. $g = g_\mathit{eq}(\bm{\chi})$ where
\begin{equation}
	g_\mathit{eq}(\bm{\chi}) = \left(\phi_2' / \phi_1'\right)^{-1}\left(-W(\bm{\chi})\right) .
	\label{GEq}
\end{equation}
In this case, the internal variable is instantaneously modified when the strain varies: no slow dynamics occurs.

The previous choice ensures that the Clausius-Duhem inequality is satisfied, independently of the sign of $\dot{g}$. Indeed, with the assumption (\ref{RelG}), the dissipation per unit volume in the material (\ref{Thermo2DissF}) is
\begin{equation}
0 \,\leqslant\, \mathscr{D} =
\left\lbrace
{\addtolength{\jot}{0.4em}
	\begin{aligned}
	& 0 & &\text{if } \tau = 0\, , \\
	&\frac{\rho}{\rho_0}\frac{\left(\phi'_1\,W + \phi'_2\right)^2}{\tau_1} \quad
	& &\text{if } \tau > 0\, .
	\end{aligned}}\right.
\label{Dissipation}
\end{equation}
If $\tau=0$ or $\tau\rightarrow{+\infty}$, then no dissipation occurs: the thermodynamic process is reversible. If $0<\tau<{+\infty}$, the thermodynamic process is irreversible, which is the origin of hysteresis curves under a dynamic loading.

The effect of $g$ on the stress (\ref{ConstitutiveEq}) is specified through $\phi_1$. If $\phi_1(g)=1$ for all $g$, then no stress softening occurs. Indeed, classical hyperelasticity is recovered. If $\phi_1(g)=0$ for all $g$, then the stress does not depend on the strain any more: the material is destroyed. For the physical relevance of the constitutive law (\ref{ConstitutiveEq}), we assume that $\phi_1 > 0$. Moreover, we assume that $g=0$ entails no stress softening: $\phi_1(0) = 1$. A natural choice satisfying these requirements is
\begin{equation}
\phi_1(g) = 1 - g\, ,
\label{SofteningLin}
\end{equation}
where $g<1$.

We require that $g=0$ is an equilibrium point (\ref{GEq}) if no strain is applied. Hence, one must have $\phi_2'(0) = 0$. If the softening function (\ref{SofteningLin}) is chosen, the convexity of $\phi_2$ ensures that the equilibrium point (\ref{GEq}) is unique. Simple choices for $\phi_2$ are
{\addtolength{\jot}{0.4em}
	\begin{align}
	&\phi_2(g) = \frac{1}{2}\gamma g^2 \, ,\label{EnergyDefectsPow}\\
	&\phi_2(g) = -\frac{1}{2} \gamma \ln(1 - g^2) \, ,\label{EnergyDefectsLog}
	\end{align}\hspace{-0.4em}}
where $\gamma > 0$ is an energy per unit volume. The choice (\ref{SofteningLin})-(\ref{EnergyDefectsLog}) ensures that $g$ is bounded by 1. In the vicinity of $g=0$, both expressions (\ref{EnergyDefectsPow}) and (\ref{EnergyDefectsLog}) are equivalent.

To summarize, the equations of motion in Lagrangian coordinates are
\begin{equation}
	\left\lbrace
	{\addtolength{\jot}{0.4em}
		\begin{aligned}
			&\dot{\bm{F}} = \mathbf{grad}\,\bm{v} \, ,\\
			&\rho_0\,\dot{\bm{v}} = \mathbf{div}\left(\phi_1 \det(\bm{F})\, \bar{\bm\sigma}\cdot\bm{F}^{-\top}\right) ,\\
			&{-\tau_1}\, \dot{g} = \phi'_1\,W + \phi'_2 \, ,
	\end{aligned}}
	\right.
	\label{SystF}
\end{equation}
where $\phi_1$ and $\phi_2$ are specified by (\ref{SofteningLin}) and (\ref{EnergyDefectsPow})-(\ref{EnergyDefectsLog}), respectively.
The expression of the hyperelastic stress $\bar{\bm \sigma}$ is specified by (\ref{StressF}) if the right Cauchy-Green tensor $\bm{\chi} = \bm{C}$ is used. Otherwise, elementary tensor algebra yields the expression of the constitutive law.

In the next section, a few cases are detailed: the isotropic case, the case of infinitesimal strain and the case of uniaxial strain.

\subsection{Particular cases}\label{sec:Particular}

\paragraph{Isotropic case.} The dependence to $\bm C$ of the internal energy can be replaced by a dependence to the invariants
\begin{equation}
{\addtolength{\jot}{0.4em}
	\begin{aligned}
	&C_\mathrm{I} = \mathrm{tr}({\bm C}) \, ,\\
	&C_\mathrm{II} = \frac{1}{2}\left(\mathrm{tr}({\bm C})^2 - \mathrm{tr}({\bm C}^2)\right) ,\\
	&C_\mathrm{III} = \det ({\bm C}) \, .
	\end{aligned}}
\label{Invariants}
\end{equation}
In particular, the conservation of mass (\ref{ConsMassF}) rewrites as $\rho_0/\rho = \sqrt{C_\mathrm{III}}$. The hyperelastic stress $\bar{\bm \sigma}$ satisfies (\ref{StressF}), where
\begin{equation}
\frac{\partial W}{\partial\bm{C}} = \frac{\partial W}{\partial{C_\mathrm{I}}} \frac{\partial C_\mathrm{I}}{\partial\bm{C}}
+ \frac{\partial W}{\partial{C_\mathrm{II}}} \frac{\partial C_\mathrm{II}}{\partial\bm{C}}
+ \frac{\partial W}{\partial{C_\mathrm{III}}} \frac{\partial C_\mathrm{III}}{\partial\bm{C}}\, ,
\label{dWdC}
\end{equation}
with the tensor derivatives~\cite{holzapfel00}
\begin{equation}
{\addtolength{\jot}{0.4em}
	\begin{aligned}
	&\frac{\partial C_\mathrm{I}}{\partial \bm{C}} = \bm{G}\, ,\\
	&\frac{\partial C_\mathrm{II}}{\partial \bm{C}} = C_\mathrm{I}\bm{G} - \bm{C}\, ,\\
	&\frac{\partial C_\mathrm{III}}{\partial \bm{C}} = C_\mathrm{II} \bm{G} - C_\mathrm{I}\bm{C} + \bm{C}^2 \, .
	\end{aligned}}
\end{equation}
Thus, the following substitution
\begin{equation}
\frac{\partial W}{\partial\bm{C}} = \left(\frac{\partial W}{\partial{C_\mathrm{I}}} + C_\mathrm{I}\frac{\partial W}{\partial{C_\mathrm{II}}} + C_\mathrm{II}\frac{\partial W}{\partial{C_\mathrm{III}}}\right)\! \bm{G}
- \left(\frac{\partial W}{\partial{C_\mathrm{II}}} + C_\mathrm{I}\frac{\partial W}{\partial{C_\mathrm{III}}}\right)\! \bm{C}
+ \frac{\partial W}{\partial{C_\mathrm{III}}} \bm{C}^2
\end{equation}
can be made in equation (\ref{StressF}). In the literature, several strain energy density functions can be found. In terms of the invariants of $\bm C$, a classical example is the compressible Mooney-Rivlin model~\cite{rivlin48}
\begin{equation}
W = \mathfrak{c}_{1}\, (C_\mathrm{I}\, {C_\mathrm{III}}^{-1/3} - 3) + \mathfrak{c}_{2}\, (C_\mathrm{II}\, {C_\mathrm{III}}^{-2/3} - 3) + \mathfrak{d}_{1}\, ({C_\mathrm{III}}^{1/2} - 1)^2\, ,
\label{Mooney-Rivlin}
\end{equation}
where $(\mathfrak{c}_{1},\mathfrak{c}_{2},\mathfrak{d}_1)$ are material parameters. This hyperelastic model (\ref{Mooney-Rivlin}) is classically used in mechanics of elastomers.

Sometimes, the strain energy density function is expressed in terms of the Green-Lagrange strain tensor $\bm{E} = \frac{1}{2}(\bm{C} - \bm{G})$ (see e.g. \cite{norris98}). An example of strain energy density in terms of the invariants of $\bm E$ is the Murnaghan's law~\cite{murnaghan37}
\begin{equation}
W = \frac{\lambda + 2\mu}{2}\, {E_\mathrm{I}}^2 - 2\mu\, E_\mathrm{II} + \frac{\mathfrak{l} + 2\mathfrak{m}}{3}\, {E_\mathrm{I}}^3 - 2\mathfrak{m}\,E_\mathrm{I}\,E_\mathrm{II} + \mathfrak{n}\,E_\mathrm{III} \, ,
\label{Murnaghan}
\end{equation}
where $(\lambda,\mu)$ are the Lamé parameters and $(\mathfrak{l},\mathfrak{m},\mathfrak{n})$ are the Murnaghan coefficients. The latter are third-order elastic constants. The hyperelastic model (\ref{Murnaghan}) is widely used in the community of nondestructive testing~\cite{johnson96a,payan09}. For conversions, one has the following relations between the invariants of $\bm{E}$ and $\bm{C}$:
\begin{equation}
{\addtolength{\jot}{0.4em}
	\begin{aligned}
	&{E}_\mathrm{I} = \frac{1}{2} \left({C}_\mathrm{I} - 3\right) &\quad & {C}_\mathrm{I} = 3 + 2 {E}_\mathrm{I}\, ,\\
	&{E}_\mathrm{II} = \frac{1}{4} \left(3 - 2 {C}_\mathrm{I} + {C}_\mathrm{II}\right) &\Leftrightarrow\quad & {C}_\mathrm{II} = 3 + 4 E_\mathrm{I} + 4 E_\mathrm{II}\, ,\\
	&{E}_\mathrm{III} = \frac{1}{8} \left({C}_\mathrm{I} - {C}_\mathrm{II} + {C}_\mathrm{III} - 1\right) &\quad & {C}_\mathrm{III} = 1 + 2 E_\mathrm{I} + 4 E_\mathrm{II} + 8 E_\mathrm{III}\, .
	\end{aligned}}
\label{InvariantsE}
\end{equation}

\paragraph{Infinitesimal strain.}
The Green-Lagrange strain tensor is linearised with respect to the displacement:
\begin{equation}
	{\bm E} \simeq \frac{1}{2} \left(\mathbf{grad}\,\bm{u} + \mathbf{grad}^\top\bm{u}\right) = \bm{\varepsilon}\, ,
	\label{HPP}
\end{equation}
where $\bm{\varepsilon} = \frac{1}{2}(\bm{F} + \bm{F}^\top) - \bm{G}$ is the infinitesimal strain tensor. Murnaghan's law is used and the expression of the first Piola-Kirchhoff stress tensor $\det(\bm{F})\,\bar{\bm\sigma}\cdot\bm{F}^{-\top}\!$ is linearised with respect to the coordinates of $\bm{\varepsilon}$:
\begin{equation}
	\det(\bm{F})\, \bar{\bm\sigma}\cdot\bm{F}^{-\top}\! \simeq \bar{\bm\sigma} \simeq \frac{\partial W}{\partial {\bm \varepsilon}} \, .
	\label{SigmaHPP}
\end{equation}
The equations of motion (\ref{SystF}) reduce to
\begin{equation}
\left\lbrace
{\addtolength{\jot}{0.4em}
	\begin{aligned}
	&\dot{\bm{\varepsilon}} = \frac{1}{2}\left(\mathbf{grad}\,\bm{v} + \mathbf{grad}^\top\bm{v}\right) ,\\
	&\rho_0\,\dot{\bm{v}} = \mathbf{div}\left(\phi_1\, \bar{\bm\sigma}\right) ,\\
	&{-\tau_1}\, \dot{g} = \phi'_1\,W + \phi'_2 \, ,
	\end{aligned}}
\right.
\label{SystEps}
\end{equation}
which is nonlinear due to the slow dynamics. Classical elastodynamics are recovered if $\tau_1 \rightarrow {+\infty}$ in (\ref{SystEps}).

\paragraph{Uniaxial strain.}
In this case, only one component of the displacement field remains. The corresponding coordinate $u$ is assumed to be invariant with respect to the other coordinates. Thus, the equations of motion (\ref{SystF}) write now as a $3\times 3$ differential system:
\begin{equation}
\left\lbrace
{\addtolength{\jot}{0.4em}
	\begin{aligned}
	&\dot{\varepsilon} = \partial_x v \, , \\
	&\rho_0\, \dot{v} = \partial_x (\phi_1\, \bar{\sigma})\, , \\
	&{-\tau_1}\, \dot{g} = \phi'_1\,W + \phi'_2 \, ,
	\end{aligned}}
\right.
\label{SystHPP}
\end{equation}
where $\partial_x$ is the space derivative, $\varepsilon = \partial_x u$ is the strain and $v = \dot{u}$ is the particle velocity. The hyperelastic stress satisfies $\bar{\sigma} = W'(\varepsilon)$, where $W'$ is the derivative of the strain energy density function~\cite{drumheller98}.

The functions $\phi_1$ and $\phi_2$ are specified by (\ref{SofteningLin}) and (\ref{EnergyDefectsPow})-(\ref{EnergyDefectsLog}), respectively. An example of strain energy density function is given by Landau's law~\cite{meurer02,li15,lombardWM15}:
\begin{equation}
	W = \left(\frac{1}{2} - \frac{\beta}{3}\varepsilon - \frac{\delta}{4}\varepsilon^2\right)\! E \varepsilon^2\, ,
	\label{EnergyElastLandau}
\end{equation}
where $E$ is the Young's modulus and $(\beta,\delta)$ are higher-order elastic constants. When $\beta$ and $\delta$ are zero, Hooke's law of linear elasticity
\begin{equation}
	W = \frac{1}{2} E\varepsilon^2
	\label{EnergyElastHooke}
\end{equation}
is recovered.

The relationship between Murnaghan's law (\ref{Murnaghan}) and Landau's law (\ref{EnergyElastLandau}) is the following. If the uniaxial approximation is made, then $E_\mathrm{I} = \varepsilon \left(1+\frac{1}{2} \varepsilon\right)$ and $E_\mathrm{II} = E_\mathrm{III} = 0$ in (\ref{Murnaghan}). A polynomial expression of the strain energy density function with respect to $\varepsilon$ is obtained,
\begin{equation}
	W = \left( \frac{1}{2} + \left(\frac{1}{2} + \frac{\vartheta}{3}\right) \varepsilon + \left(\frac{1}{8} + \frac{\vartheta}{2}\right) \varepsilon^2 \right) \left(\lambda + 2\mu\right) \varepsilon^2 + \mathcal{O}(\varepsilon^5) \, ,
	\label{EnergyMurnaghanLandau}
\end{equation}
where $\vartheta=(\mathfrak{l} + 2\mathfrak{m})/(\lambda + 2 \mu)$. By identification with Landau's law (\ref{EnergyElastLandau}), the parameters $(E,\beta,\delta)$ can be expressed in terms of the Lamé and Murnaghan parameters:
\begin{equation}
	E = \lambda + 2\mu\, , \qquad
	\beta = -\frac{3}{2} - \vartheta\, ,\qquad
	\delta = -\frac{1}{2} - 2\vartheta\, .
	\label{ParamMurnaghanLandau}
\end{equation}
A similar calculus can be performed with the Mooney-Rivlin model (\ref{Mooney-Rivlin}).


\section{Analysis of the model}\label{sec:AnalysisModel}


\subsection{Analytical results}

From now on, the softening function (\ref{SofteningLin}) is used. If a strain step is applied locally, then $g$ is driven by (\ref{EvolG}), where the strain energy $W$ is a constant. With the quadratic expression (\ref{EnergyDefectsPow}) of $\phi_2$, the internal variable $g$ evolves exponentially in time towards $g_\mathit{eq}(\bm{\chi})$, which is defined in (\ref{GEq}). The corresponding relaxation time is $\tau_\gamma = \tau_1/\gamma$.

Now, the case of uniaxial strain is considered. A sinusoidal strain with frequency $f_c = \omega_c/2\pi = 10$~kHz and amplitude $V$ is applied locally. With the quadratic expression  (\ref{EnergyDefectsPow}) of $\phi_2$, the evolution equation (\ref{EvolG}) writes
\begin{equation}
{\addtolength{\jot}{0.4em}
	\begin{aligned}
	\dot{g}(t) + \frac{g(t)}{\tau_\gamma} & = \frac{1}{\tau_{EV}} \frac{W(V\!\sin(\omega_c t))}{E V^2} \, ,\\
	& = \frac{1}{\tau_{EV}} \left( \frac{a_0}{2} + \sum_{n=1}^{\infty} a_n\cos(n \omega_c t) + b_n\sin(n \omega_c t) \right) ,
	\end{aligned}}
\label{GEDO}
\end{equation}
where $\tau_{EV} = \tau_1/(EV^2)$ is a time constant and $(a_n, b_n)$ are the Fourier coefficients of the normalized strain energy $W/(EV^2)$.

The solution of the ordinary differential equation (\ref{GEDO}) is
\begin{equation}
{\addtolength{\jot}{0.4em}
	\begin{aligned}
	g(t) & = \left(g(0) - \frac{\tau_\gamma}{\tau_{EV}}\!\left(\frac{a_0}{2} + \sum_{n=1}^{\infty} \frac{a_n - n\omega_c\tau_\gamma b_n}{1 + (n\omega_c\tau_\gamma)^2}\right)\right)\exp(-t/\tau_\gamma) \\
	& + \frac{\tau_\gamma}{\tau_{EV}}\!\left(\frac{a_0}{2} + \sum_{n=1}^{\infty} \frac{a_n - n\omega_c\tau_\gamma b_n}{1 + (n\omega_c\tau_\gamma)^2}\cos(n \omega_c t) + \frac{b_n + n\omega_c\tau_\gamma a_n}{1 + (n\omega_c\tau_\gamma)^2}\sin(n \omega_c t) \right) .
	\end{aligned}}
\label{GEDOSol}
\end{equation}
The first term in (\ref{GEDOSol}) decreases exponentially in time with constant $\tau_\gamma$. The second term is the steady-state term, which oscillates at the frequency $f_c$ around its average value
\begin{equation}
	\left\langle g \right\rangle_{t\gg\tau_\gamma} = \frac{\tau_\gamma}{\tau_{EV}}\frac{a_0}{2}\, ,
	\label{GMean}
\end{equation}
where $\tau_\gamma/\tau_{EV} = EV^2/\gamma$.

In the case of Landau's law (\ref{EnergyElastLandau}), the nonzero Fourier coefficients are given in table~\ref{tab:FourierLandau}. At small strain amplitudes, $\beta V \ll 1$ and $\delta V^2 \ll 1$, the high-order terms in table~\ref{tab:FourierLandau} can be neglected. Thus, the case of Hooke's law (\ref{EnergyElastHooke}) is recovered, where $a_0=1/2$ and $a_2=-1/4$ are the only nonzero Fourier coefficients. In particular, the value of the average of $g$ (\ref{GMean}) is very close to the value obtained in the case of Hooke's law:
\begin{equation}
	\left\langle g \right\rangle_{t\gg\tau_\gamma} = \frac{E}{4\gamma} V^2 + \mathcal{O}(V^4) \, .
	\label{GMeanLandau}
\end{equation}
From a practical point of view, if the Young's modulus $E$ is known and the constants $\tau_\gamma$ and $\left\langle g \right\rangle_{t\gg\tau_\gamma}$ are deduced from measurements at small sinusoidal loadings, then the parameters $\tau$ and $\gamma$ of the model can be estimated.

\begin{table}
	\caption{Nonzero Fourier coefficients (\ref{GEDO}) in the case of Landau's law (\ref{EnergyElastLandau}).\label{tab:FourierLandau}}
	
	\centering
	
	{\renewcommand{\arraystretch}{1.4}
		\renewcommand{\tabcolsep}{0.25cm}
		\begin{tabular}{cccccc}
			\toprule
			$n$ & 0 & 1 & 2 & 3 & 4 \\
			\midrule
			$a_n$ & $\frac{1}{2} - \frac{3}{16}\delta V^2$ & & $-\frac{1}{4} + \frac{1}{8}\delta V^2$ & & $-\frac{1}{32}\delta V^2$ \\
			$b_n$ & & $-\frac{1}{4}\beta V$ & & $\frac{1}{12}\beta V$ & \\
			\bottomrule
	\end{tabular}}
\end{table}

In the 1D case (\ref{SystHPP}), the speed of sound is
\begin{equation}
	c = \sqrt{\frac{1}{\rho_0}\frac{\partial\sigma}{\partial\varepsilon}} =  \sqrt{\frac{(1-g)\, \bar{\sigma}'(\varepsilon)}{\rho_0}}\, .
	\label{GSpeedSound}
\end{equation}
If the material is linear-elastic without slow dynamics, the speed of sound reduces to $c_0 = \sqrt{E/\rho_0}$. It is easier for the analysis to introduce the elastic modulus $M = \rho_0\, c^2$ and its variation
\begin{equation}
	\frac{\Delta M}{M} = \frac{\rho_0\, c^2 - E}{E} = (1-g)\,\frac{\bar{\sigma}'(\varepsilon)}{E} - 1 \, .
	\label{GMVar}
\end{equation}
On figure~\ref{fig:RiviereSoftening}, the experimental variation in speed of sound $\Delta c/c = (c - c_0)/c_0$ is represented instead.

When Landau's law (\ref{EnergyElastLandau}) is used, the variation in elastic modulus is
\begin{equation}
	\frac{\Delta M}{M} = (1-g)\,(1 - 2\beta\varepsilon - 3\delta\varepsilon^2) - 1\, ,
	\label{GMVarLandau}
\end{equation}
which reduces to $-g$ if $\beta$ and $\delta$ equal zero. The average of $\Delta M/M$ over a period of forcing is deduced from (\ref{GEDOSol}) and (\ref{GMVarLandau}):
\begin{equation}
	\left\langle \frac{\Delta M}{M} \right\rangle_{t\gg \tau_\gamma}
	= -\frac{E + 6\delta\gamma}{4\gamma} V^2 + \mathcal{O}(V^4) \, .
	\label{MMeanHooke}
\end{equation}
The diminution of the elastic modulus with the square of the strain amplitude is similar to the \emph{Payne effect} in filled rubber~\cite{payne62}.

\begin{table}
	\caption{Physical parameters.\label{tab:ParamElast}}
	\centering
	{\renewcommand{\arraystretch}{1.2}
		\renewcommand{\tabcolsep}{0.5cm}
		\begin{tabular}{llllll}
			\toprule
			$\rho_0$ (kg.m$^{-3}$) & $E$ (GPa) & $\gamma$~(J.m\textsuperscript{-3}) & $\tau$ (s) \\
			$2600$ & $10$ & $20$ & $7\times 10^{-3}$ \\
			\bottomrule
	\end{tabular}}
\end{table}

\begin{figure}
	\centering
	
	(a)
	
	\includegraphics{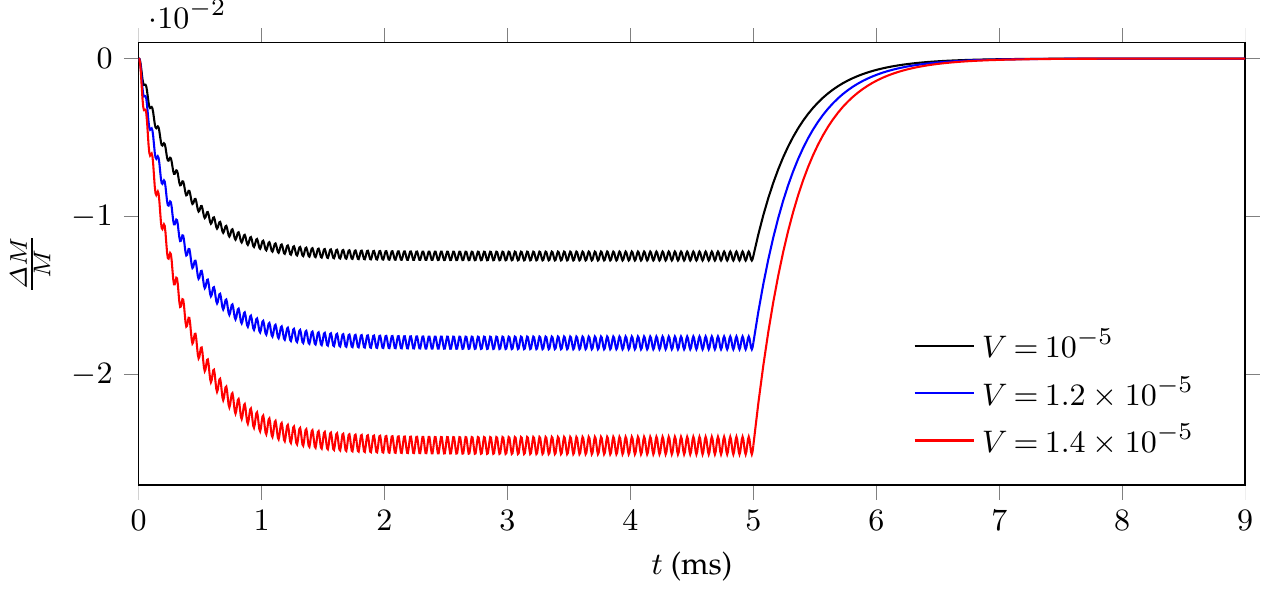}
	\vspace{0.5em}
	
	\begin{minipage}{0.49\textwidth}
		\centering
		
		(b)
		
		\includegraphics{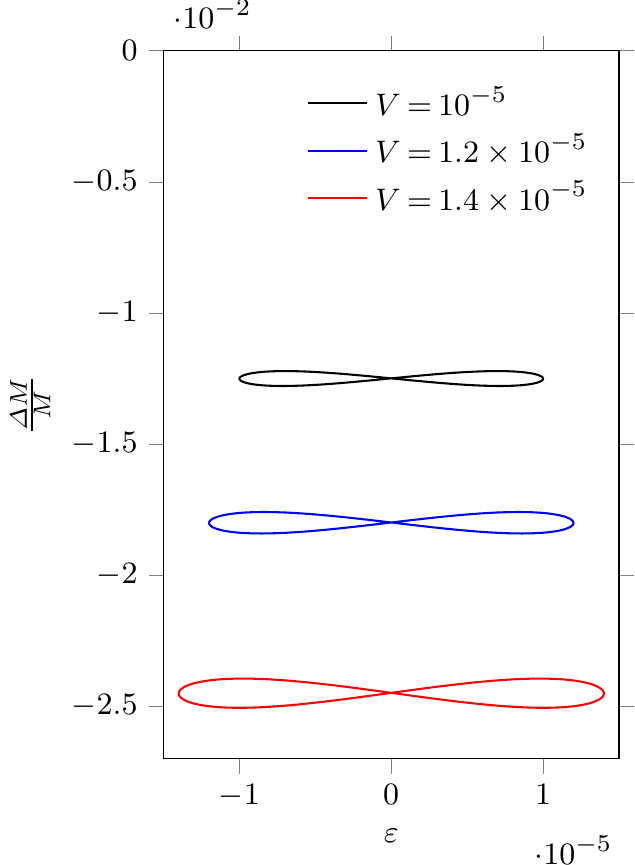}
	\end{minipage}
	\begin{minipage}{0.49\textwidth}
		\centering
		
		(c)
		
		\includegraphics{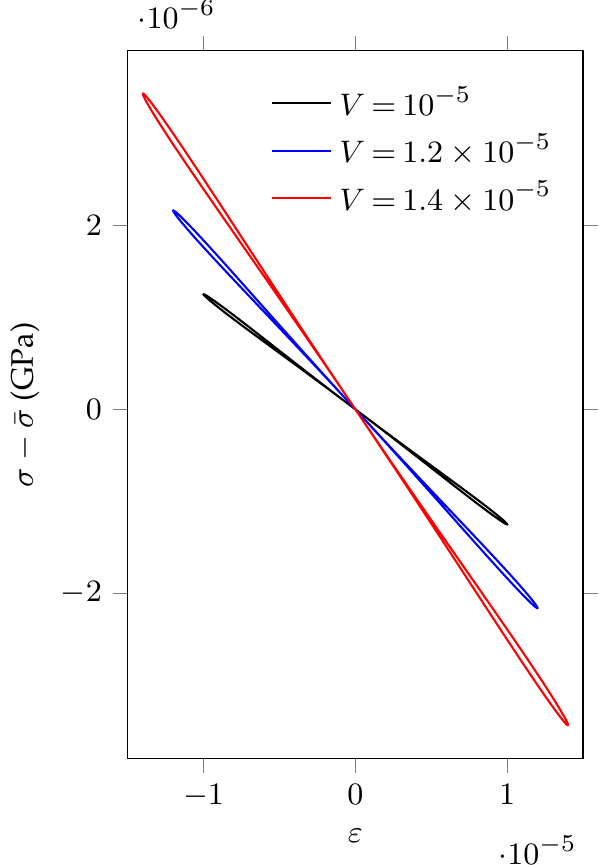}
	\end{minipage}
	
	\caption{Analytical computation in the case of Hooke's law. (a) Evolution of the relative variation in elastic modulus $\Delta M/M = -g$ with respect to its initial value, when a sinusoidal strain $\varepsilon = V \sin(\omega_c t)$ is applied until $t = 5$~ms (\ref{GEDOSol}). (b) Hysteresis curves $\Delta M/M$ versus $\varepsilon$ in steady state ($\tau_\gamma \ll t < 5$~ms); (c) effect of hysteresis on the stress-strain relationship, where the stress $\sigma-\bar{\sigma}$ is represented with respect to $\varepsilon$.\label{fig:EvolG}}
\end{figure}

On figure~\ref{fig:EvolG}-(a), $\Delta M/M$ is represented up to $t=5$~ms in the case of Hooke's law (\ref{EnergyElastHooke}) with the parameters from table~\ref{tab:ParamElast}. In this softening phase, $\Delta M/M$ decreases and reaches the steady state. At $t=5$~ms, the excitation is stopped. Thus, $\tau_{EV}$ goes to infinity in (\ref{GEDOSol}). During the recovery, $\Delta M/M$ increases exponentially in time towards zero with time constant $\tau_\gamma  = 0.35$~ms.

Figures~\ref{fig:EvolG}-(b) and \ref{fig:EvolG}-(c) show the steady-state solution. On figure~\ref{fig:EvolG}-(b), $\Delta M/M$ is represented with respect to the strain for several forcing amplitudes, according to equation (\ref{GMVarLandau}) with $\beta=\delta=0$. A hysteretic behaviour caused by the dissipation is observed. Figure~\ref{fig:EvolG}-(c) is an alternative representation of the phenomenon for several strain amplitudes. Here, the effect of increasing strain levels on the stress-strain relationship is outlined.

\begin{figure}
	\begin{minipage}{0.49\textwidth}
		\centering
		(a)
		
		\includegraphics{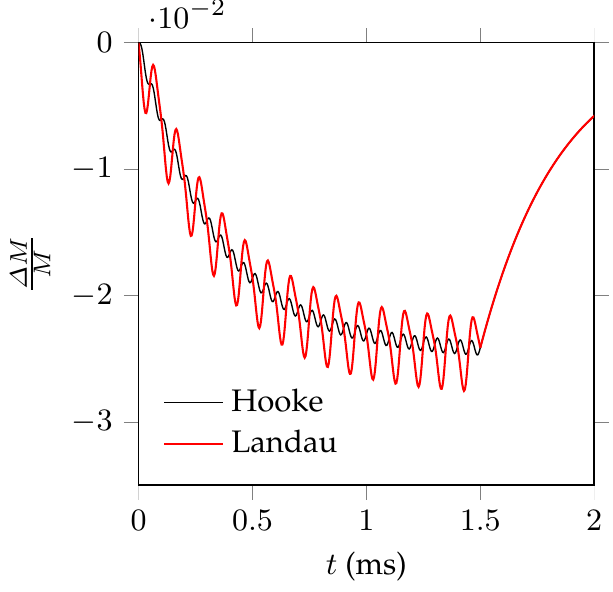}
	\end{minipage}
	\begin{minipage}{0.49\textwidth}
		\centering
		(b)
		
		\includegraphics{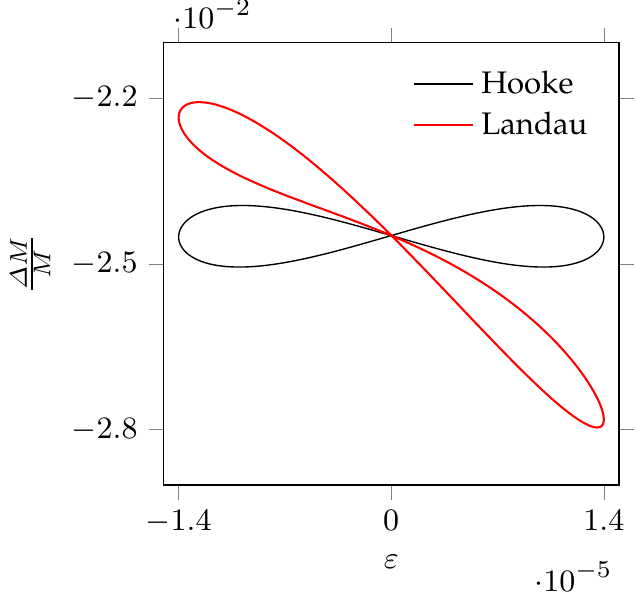}
	\end{minipage}
	
	\caption{Comparison of the analytical computations in the cases of Hooke's law and Landau's law. (a) Evolution of the variation in elastic modulus $\Delta M/M$ when a sinusoidal strain with amplitude $V = 1.4\times 10^{-5}$ is applied until $t = 1.5$~ms (\ref{GEDOSol})-(\ref{GMVarLandau}). (b) Hysteresis curves $\Delta M/M$ versus $\varepsilon$ in steady state.\label{fig:EvolGLandau}}
\end{figure}

On figure~\ref{fig:EvolGLandau}, the behaviour of our model with Landau's law (\ref{EnergyElastLandau}) and $\lbrace\beta = 10^2, \delta = 10^6\rbrace$ is compared to the previous case of Hooke's law (\ref{EnergyElastHooke}). At strain amplitudes $V \approx 10^{-5}$, the contribution of $\beta$ and $\delta$ in the Fourier coefficients is not significant (table~\ref{tab:FourierLandau}). On figure~\ref{fig:EvolGLandau}-(a), the softening phases are compared. Figure~\ref{fig:EvolGLandau}-(b) represents the hysteresis curves. More important variations of $\Delta M/M$ are observed in the case of Landau's law, as well as a loss of symmetry in the hysteresis curves. These phenomena are due to the dependence (\ref{GMVarLandau}) of $\Delta M/M$ with the strain, when $\beta$ and $\delta$ are nonzero.

Supplementary analytical results can be obtained in the case of Hooke's law (\ref{EnergyElastHooke}). In this case, the variation in elastic modulus (\ref{GMVarLandau}) is $\Delta M/M = -g$, and the only nonzero Fourier coefficients in table~\ref{tab:FourierLandau} are $a_0$ and $a_2$. The surface area of the hysteresis loops in figure~\ref{fig:EvolG}-(b) is
\begin{equation}
{\addtolength{\jot}{0.4em}
	\begin{aligned}
	S_{\bowtie} = \frac{8}{3} \frac{\tau_\gamma}{\tau_{EV}} \frac{2\omega_c \tau_\gamma}{1 + (2\omega_c\tau_\gamma)^2} V
	= \frac{4}{3} \frac{\omega_c \tau_1}{\gamma^2 + (2\omega_c\tau_1)^2} E V^3 \, ,
	\end{aligned}}
\label{SButt}
\end{equation}
which vanishes at high $\gamma$, low and high frequency $f_c$, and low and high $\tau$. The maximum value reached by the steady-state solution is
\begin{equation}
	g_\mathit{max} = \frac{\tau_\gamma}{\tau_{EV}} \left(1 + \frac{1}{\sqrt{1 + (2\omega_c\tau_\gamma)^2}}\right) .
	\label{maxG}
\end{equation}
The strain amplitude $V_\mathit{max}$ for which the material is destroyed satisfies $g_\mathit{max} = 1$:
\begin{equation}
	V_\mathit{max} = \sqrt{\frac{2\gamma}{E}}\, \sqrt{ \frac{2 \sqrt{\gamma^2 + (2\omega_c\tau_1)^2}}{\gamma + \sqrt{\gamma^2 + (2\omega_c\tau_1)^2}} } \, .
	\label{maxV}
\end{equation}
In the present configuration, $V_\mathit{max} \approx 8.8\times 10^{-5}$. Thus, if the quadratic expression (\ref{EnergyDefectsPow}) of the storage energy $\phi_2$ is chosen, the model is only valid for small strains. In the case of the logarithmic expression (\ref{EnergyDefectsLog}) of $\phi_2$, no strain limit is imposed by the slow dynamics.


\subsection{Properties}

\paragraph{Internal energy.}

According to the equation (\ref{ConstitutiveE}), the internal energy per unit volume is separated into two terms. One term corresponds to the strain energy $\phi_1 W$, the other term corresponds to the storage energy $\phi_2$. When $g=0$, the internal energy is only elastic. As $g$ increases at constant strain, the strain energy decreases and the storage energy increases. Therefore, the internal energy is transferred from the strain to $\phi_2$ when $g$ increases, and inversely.

Let us assume that $\tau = 0$. The internal variable satisfies $g = g_\mathit{eq}(\bm{\chi})$ (\ref{GEq}). With the quadratic expression (\ref{EnergyDefectsPow}) of $\phi_2$, the internal variable is equal to
\begin{equation}
	g_\mathit{eq}(\bm{\chi}) = \frac{W(\bm{\chi})}{\gamma}\, .
	\label{GEqQuad}
\end{equation}
The value $g = 1$, which corresponds to a destructed material, is reached for strain energies $W\geqslant\gamma$. In the case of Hooke's law (\ref{EnergyElastHooke}) with the parameters from table~\ref{tab:ParamElast}, the maximum admissible strain is $\sqrt{2\gamma/E} \approx 6.3\times 10^{-5}$. This value is recovered by setting $\tau_1=0$ in equation (\ref{maxV}). The logarithmic expression (\ref{EnergyDefectsLog}) of $\phi_2$ yields
\begin{equation}
g_\mathit{eq}(\bm{\chi}) = \frac{2 W(\bm{\chi})}{\gamma + \sqrt{\gamma^2 + 4 W(\bm{\chi})^2}} \, ,
\label{GEqLog}
\end{equation}
which is always between zero and one. Therefore, there is no strain limit in this case.

\begin{figure}
	\centering
	
	\includegraphics{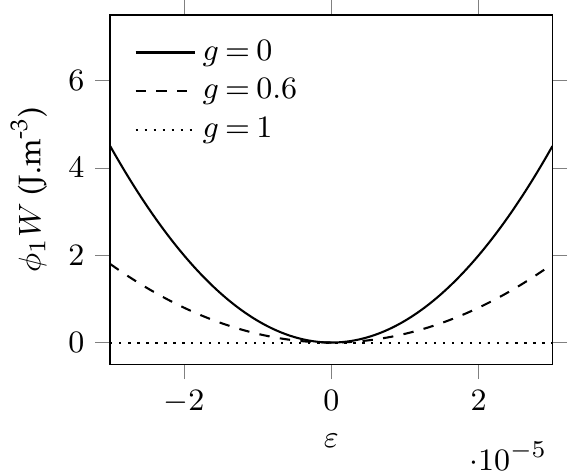}

	\caption{
		Sketch of the strain energy per unit volume $\phi_1 W = (1-g) \frac{1}{2}E\varepsilon^2$ with respect to the strain $\varepsilon$, for several values of the internal variable $g$.
		\label{fig:EnergyEps}}
\end{figure}

Figure~\ref{fig:EnergyEps} represents the strain energy per unit volume $\phi_1 W$ when the geometry is 1D. The strain energy density function is issued from Hooke's law (\ref{EnergyElastHooke}) and the softening function (\ref{SofteningLin}) is used (parameters from table~\ref{tab:ParamElast}). One can observe that the strain energy decreases as $g$ increases. If $g=1$, the strain energy does not depend on the strain anymore, which illustrates the destruction of the material.

\begin{figure}
	\begin{minipage}{0.47\textwidth}
		\centering
		(a)
		
		\includegraphics{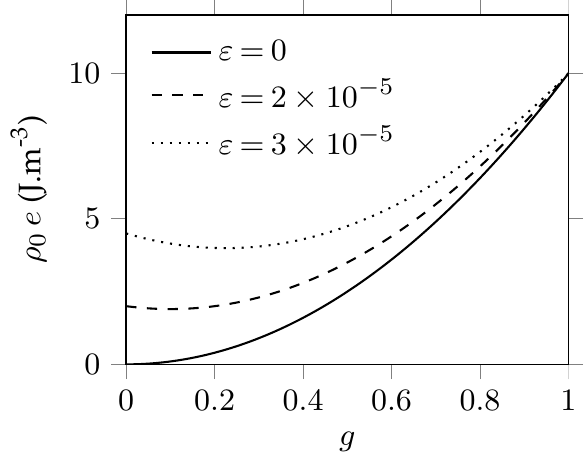}
	\end{minipage}
	\hfill
	\begin{minipage}{0.47\textwidth}
		\centering
		(b)
		
		\includegraphics{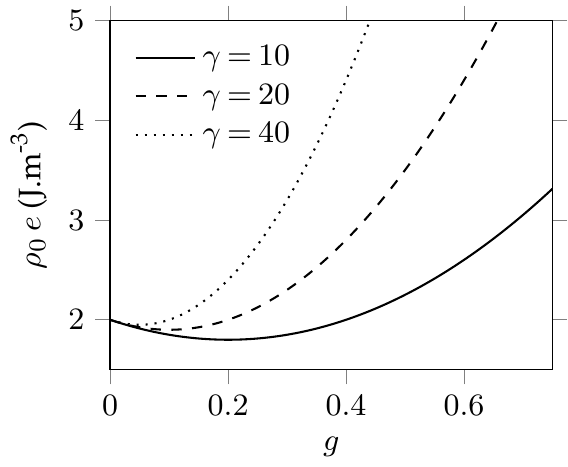}
	\end{minipage}
	
	\caption{Sketch of the internal energy per unit volume $\rho_0 e$ with respect to $g$ (\ref{ConstitutiveE}). It is represented (a) for several values of $\varepsilon$ when $\gamma=20$~J.m\textsuperscript{-3}; (b) for several values of $\gamma$ (J.m\textsuperscript{-3}) when $\varepsilon = 2\times 10^{-5}$.\label{fig:EnergyG}}
\end{figure}

On figures \ref{fig:EnergyG}-(a) and \ref{fig:EnergyG}-(b), the internal energy is represented with respect to $g$, where the quadratic expression (\ref{EnergyDefectsPow}) of the storage energy $\phi_2$ is used. The values of $g_\mathit{eq}$ correspond to the abscissas of the local minima of the curves (\ref{GEqQuad}). On figure~\ref{fig:EnergyG}-(a), one can observe an increase in $g_\mathit{eq}$ when the strain increases. No asymptote avoids to reach the value $g=1$, which destroys the material. On figure~\ref{fig:EnergyG}-(b), one can observe an increase in $g_\mathit{eq}$ when $\gamma$ decreases. Again, no asymptote avoids to reach the value $g=1$, which destroys the material.

\paragraph{Dissipation.}

\begin{figure}
	\centering
	\includegraphics{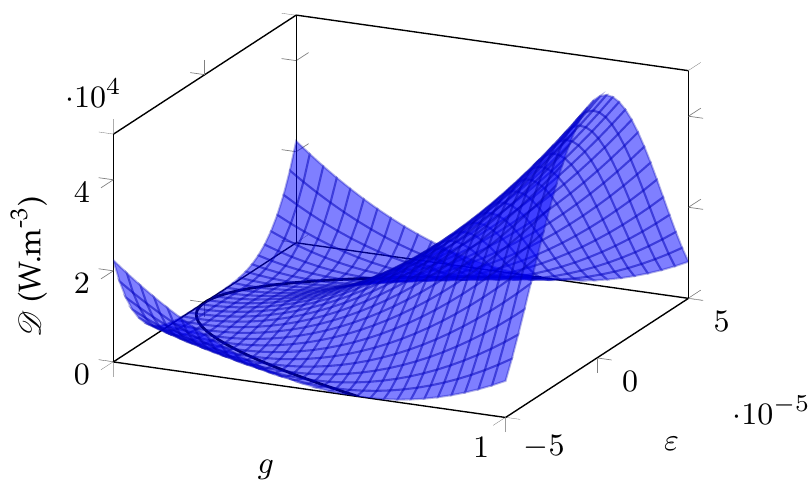}
	
	\caption{View of the dissipation $\mathscr{D}$ in $\varepsilon$-$g$ coordinates (\ref{Dissipation}). The black line marks the curve $g=g_\mathit{eq}(\varepsilon)$, i.e. the locus $\tau=0$ where no dissipation occurs (\ref{GEq}).\label{fig:Dissipation}}
\end{figure}

In one space dimension and small strain, $\mathscr{D}$ depends on $\varepsilon$ and $g$. The dissipation per unit volume (\ref{Dissipation}) is a surface in $\varepsilon$-$g$ coordinates (figure~\ref{fig:Dissipation}). The expression of $\mathscr{D}$ is deduced from the softening function (\ref{SofteningLin}), the quadratic storage energy (\ref{EnergyDefectsPow}), Hooke's law (\ref{EnergyElastHooke}) and the conservation of mass $\rho_0/\rho = 1+\varepsilon$. This figure illustrates that the dissipation is positive, in agreement with the Clausius-Duhem inequality. Also, one can observe that no dissipation occurs if $\tau=0$, which corresponds to the curve $g = g_\mathit{eq}(\varepsilon) =  E\varepsilon^2/(2\gamma)$.

\section{Conclusion}

A new model for the dynamic behaviour of solids is proposed. The following features are common with the soft-ratchet model of Vakhnenko et al.~\cite{vakhnenko04}:
\begin{itemize}
	\item a variable $g$ describes the softening of the material;
	\item an evolution equation for $g$ with a relaxation time $\tau$ is given;
	\item a low number of extra parameters for the non-classical effects is required.
\end{itemize}
In comparison with the soft-ratchet model, several differences can be outlined:
\begin{enumerate}[(i)]
	\item the new model satisfies the second principle of thermodynamics;
	\item the new model does not require an expression for the equilibrium value $g_\mathit{eq}(\sigma)$ of $g$, but an expression of the storage energy $\phi_2(g)$;
	\item the new model generalizes naturally to higher space dimensions.
\end{enumerate}
The point (i) is a major difference (see appendix~\ref{sec:Vakhnenko}), which ensures that our model is thermodynamically relevant. As shown in section~\ref{sec:AnalysisModel}, the new model reproduces qualitatively the macroscopic behaviour of real media.

Our approach is purely phenomenological. No physical interpretation of $g$ at the microscopic scale is known. To go further, some similarities with other materials are pointed out in appendix~\ref{sec:Extensions}, in particular with filled rubber. It seems that the dynamic response of rocks is similar to the Payne effect~\cite{payne62}, and that the quasistatic response of rocks is similar to the Mullins effect~\cite{diani09,machado10}. In mechanics of elastomers, existing quasistatic models have a very similar structure to our dynamic model~\cite{ogden99,dorfmann03}. By analogy, the coupling of nonlinear viscoelasticity and heat conduction could be a key for future physical modelling (see e.g. \cite{holzapfel96b}). Lastly, from a microscopic point of view, both materials are roughly made of a matrix with particles inside. These similarities may be used for future micromechanical modelling.

Future work will be devoted to 2D and 3D numerical modelling of the nonlinear wave propagation. Since the system of partial differential equations is nonlinear, a mathematical study of the existence and the smoothness of solutions is required. Also, the computation of long-time periodic solutions will be addressed. Lastly, comparisons with real experiments should be done to validate the model.

\vskip6pt








\appendix

\section*{Appendix}

\subsection{Analogies with other models}\label{sec:Extensions}

\paragraph{Quasistatic loading of filled rubber.} In the case of a quasistatic process, equilibrium is satisfied over the transformation. This is equivalent to have $\dot{g} = 0$ in (\ref{EvolG}). The internal variable is then deduced from the strain through $g = g_\mathit{eq}(\bm{\chi})$ (\ref{GEq}). Due to the constitutive relation (\ref{ConstitutiveEq}), the stress depends explicitly on the strain. Therefore, no hysteresis occurs in the stress-strain relationship.

Pseudo-elastic models are designed to incorporate hysteresis and memory effects. Additional variables which are stored along the loading path can be used in the storage energy $\phi_2$. For example, $W_\mathit{max} = \max_t W(\bm{\chi})$ is used in~\cite{ogden99} to describe the \emph{Mullins effect}, which is observed in cyclic loading of filled rubber. An expression of the form
\begin{equation}
	\phi_2'(g) = W_\mathit{max} + \frac{2\gamma}{\sqrt{\pi}} \, \mathrm{erf}^{-1}(-g)\, ,
	\label{OgdenRoxburghPhi}
\end{equation}
is proposed in~\cite{ogden99}. From (\ref{GEq}), one deduces the expression of the internal variable
\begin{equation}
	g_\mathit{eq}(\bm{\chi}) = \mathrm{erf}\left(\frac{W_\mathit{max} - W(\bm{\chi})}{2\gamma/\sqrt{\pi}}\right) .
	\label{OgdenRoxburghG}
\end{equation}
This expression satisfies $g_\mathit{eq} = 0$ if $W(\bm{\chi}) = W_\mathit{max}$. In particular, $g_\mathit{eq} = 0$ along the primary loading path. In the case of the \emph{end-point memory} phenomenon which is observed in rocks~\cite{guyer99}, the pseudo-elastic model~\cite{ogden99} can be adapted as described in section~4 of~\cite{dorfmann03}. For further reading, a review on existing models of rubber can be found in~\cite{diani09,machado10}.

\begin{figure}
	\centering
	\includegraphics{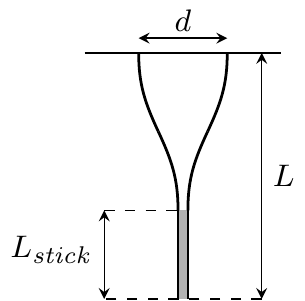}
	
	\caption{Sketch of two sticking fibers of length $L$ with initial spacing $d$, when withdrawn from a wetting liquid (grey). The height of fluid between the fibers is $L_\mathit{stick}$.}\label{fig:Fibers}
\end{figure}

\paragraph{System of wet fibers.} A formal analogy with a system of two partially-immersed fibers of length $L$ can be made (figure~\ref{fig:Fibers}). Initially, their spacing is $d$. Then, the fibers are immersed in a fluid with surface tension $\Upsilon$. When withdrawn quasi-statically, they stick together. The internal energy of this system is the sum of the bending energy in the fibers and the energy due to the surface tension of the fluid. Thus,~\cite{py07a}
\begin{equation}
\mathcal{E}_\mathit{int} = \phi_1(g)\, \mathcal{E}_\mathit{el} + \phi_2(g) \, ,
\qquad g = \frac{L_\mathit{stick}}{L} \in \left[0,1\right[ ,
\label{FibEint}
\end{equation}
where $L_\mathit{stick}$ is the wet length of the fibers. In the case of a system of cylindrical elastic fibers with radius $r$ and Young's modulus $E$, the expressions in (\ref{FibEint}) are
\begin{equation}
{\addtolength{\jot}{0.4em}
	\begin{aligned}
	&\phi_1(g) = (1-g)^{-3} \, ,\\
	&\mathcal{E}_\mathit{el} = \frac{3 E I d^2}{L^3}
	\qquad\text{with}\qquad I = \frac{\pi r^4}{4} \, ,\\
	&\phi_2(g) = {-4}\Upsilon r L \int_0^{g} \left(\theta - \left(\frac{\pi}{2} - \theta\right)\left(\frac{1}{\cos\theta} - 1\right)\right) d\zeta \, ,
	\end{aligned}}
\label{FibExp}
\end{equation}
Due to the geometry of the meniscus and the law of hydrostatics, one has
\begin{equation}
	\cos\theta = \frac{\rho_\mathrm{f}\, g_\mathrm{n} \zeta r L}{\Upsilon + \rho_\mathrm{f}\, g_\mathrm{n} \zeta r L}
	\qquad\text{with}\qquad
	\zeta = \frac{z}{L}\, ,
	\label{FibTheta}
\end{equation}
where $z$ is the altitude in the fluid, $\rho_\mathrm{f}$ is the mass density of the fluid and $g_\mathrm{n}$ is the standard gravity. A sign mistake has been found in equation (2) of~\cite{py07a}. Equations (\ref{FibExp})-(\ref{FibTheta}) are taken from equations (3)-(4) of~\cite{py07a}, where the sign is correct. Formally, the energy (\ref{FibEint}) is similar to the energy  (\ref{ConstitutiveE}).

\subsection{Limitations of the soft-ratchet model}\label{sec:Vakhnenko}

\paragraph{Thermodynamical analysis.} The soft-ratchet model is a particular case of 1D model with internal variable of state~\cite{vakhnenko04}. Thus, we carry out the thermodynamical analysis from section~\ref{sec:Model}. The soft-ratchet model introduces a concentration of activated defects $g$, which modifies the stress according to
\begin{equation}
	\sigma = (1-g)\, \bar{\sigma}(\varepsilon) \, .
	\label{ConstitutiveVakh}
\end{equation}
This constitutive law is the same as (\ref{ConstitutiveEq}) with the softening function (\ref{SofteningLin}). In one space dimension, the strain rate satisfies $D = \dot{\varepsilon}/F$, where $F = 1+\varepsilon$. The Clausius-Duhem inequality (\ref{Thermo2Diss}) rewrites as
\begin{equation}
{\addtolength{\jot}{0.4em}
	\begin{aligned}
	\mathscr{D} = \left(\sigma - \rho_0 \frac{\partial e}{\partial \varepsilon}\right) \! D - \rho \frac{\partial e}{\partial g}\, \dot{g} \geqslant 0\, ,
	\end{aligned}}
\label{Thermo2DissVakh}
\end{equation}
for all state and all evolution. Due to the constitutive law (\ref{ConstitutiveVakh}), the specific internal energy must satisfy
\begin{equation}
	\rho_0\frac{\partial e}{\partial \varepsilon} = (1-g)\, \bar{\sigma}(\varepsilon)\, .
	\label{ConstitutiveEVakh}
\end{equation}
When integrating (\ref{ConstitutiveEVakh}) with respect to the strain $\varepsilon$, an integration constant appears, which we denote by $\phi_2(g)$. Thus, the internal energy per unit volume (\ref{ConstitutiveE}) is recovered, where $W'(\varepsilon) = \bar{\sigma}(\varepsilon)$. The Clausius-Duhem inequality (\ref{Thermo2DissVakh}) implies
\begin{equation}
	\left(W(\varepsilon) - \phi'_2(g)\right) \dot{g} \geqslant 0
	\label{DissipationVakh}
\end{equation}
for all $\lbrace\varepsilon,g\rbrace$ and all $\dot g$.

In the soft-ratchet model, the evolution equation for $g$ has the form
\begin{equation}
\dot{g} = -\frac{1}{\tau}\left(g - g_\mathit{eq}(\sigma)\right) \, ,
\label{EvolGVakh}
\end{equation}
where $\tau > 0$ is a variable relaxation time and $g_\mathit{eq}(\sigma)$ is the value of $g$ at equilibrium for a given stress. Various expressions of $g_\mathit{eq}$ are proposed in the literature. In~\cite{vakhnenko04}, $g_\mathit{eq}$ reads
\begin{equation}
	g_\mathit{eq}(\sigma) = g_0\exp\!\left(\frac{\sigma}{\tilde{\sigma}}\right) ,
	\label{gEqVakh}
\end{equation}
where $\tilde{\sigma}$ is a stress and $g_0$ is the value of $g_\mathit{eq}$ at zero stress.	This expression is modified in~\cite{lombardWM15} to ensure $g_\mathit{eq} < 1$:
\begin{equation}
	g_\mathit{eq}(\sigma) = 
	\frac{1}{2}\left(1 + \tanh\!\left(\frac{\sigma}{\tilde{\sigma}} - \tanh^{-1}(1 - 2 g_0)\right)\right) .
	\label{gEqLomb}
\end{equation}
Injecting (\ref{EvolGVakh}) in (\ref{DissipationVakh}) yields the condition
\begin{equation}
	\left(\phi_2'(g) - W(\varepsilon)\right) \left(g - g_\mathit{eq}(\sigma)\right) \geqslant 0
	\qquad\text{with}\qquad
	\sigma = (1-g)\,\bar{\sigma}(\varepsilon)\, ,
	\label{Thermo2DissVakhEQ}
\end{equation}
for all $\varepsilon$ in $\left]{-1},{+\infty}\right[$ and all $g$ in $\left[ 0,1 \right]$.

In particular, (\ref{Thermo2DissVakhEQ}) must hold for all $g$ when $\varepsilon = 0$. In this case, the condition (\ref{Thermo2DissVakhEQ}) reduces to $g \geqslant g_0$ for all $g$ such that $\phi'_2(g) > 0$. We deduce that $g_0$ must be negative or zero, i.e. $g_0 = 0$. The expressions (\ref{gEqVakh})-(\ref{gEqLomb}) of $g_\mathit{eq}$ imply that $g_\mathit{eq}$ is always equal to zero, which is not physically relevant. Something must be modified in the soft-ratchet model to satisfy equation (\ref{Thermo2DissVakhEQ}). Here, we propose to seek thermodynamically admissible expressions of $g_\mathit{eq}$.

\paragraph{Modified model.} Expressions of $g_\mathit{eq}$ must be chosen carefully. The condition (\ref{Thermo2DissVakhEQ}) imposes that $\phi_2'(g) - W(\varepsilon)$ and $g - g_\mathit{eq}(\sigma)$ have the same sign. Both functions of $\varepsilon$ and $g$ are smooth. Hence, they equal zero with a change in sign or with a gradient equal to zero. Since the gradient of both functions is nonzero, it implies that $\phi_2'(g) - W(\varepsilon)$ and $g - g_\mathit{eq}(\sigma)$ equal zero for the same values of $\varepsilon$ and $g$. Combining both equalities, the condition
\begin{equation}
	\phi'_2(g_\mathit{eq}(\sigma)) = W\! \left( \bar{\sigma}^{-1}\left(\frac{\sigma}{1-g_\mathit{eq}(\sigma)}\right) \right)
	\label{CondGeqVakh}
\end{equation}
is deduced from the constitutive law (\ref{ConstitutiveVakh}). An expression of $g_\mathit{eq}$ which satisfies (\ref{CondGeqVakh}) is not necessarily thermodynamically admissible. Moreover, one can note that such an expression depends on the strain energy density $W$ and on the storage energy $\phi_2$.

Now, we examine the existence of a thermodynamically admissible expression of $g_\mathit{eq}$ in a particular case. To do so, the strain energy density from Hooke's law (\ref{EnergyElastHooke}) is chosen. We select the quadratic expression (\ref{EnergyDefectsPow}) of the storage energy $\phi_2$. The necessary condition (\ref{CondGeqVakh}) writes
\begin{equation}
	\gamma\, g_\mathit{eq}(\sigma) = \frac{1}{2} E \left(\frac{\sigma/E}{1-g_\mathit{eq}(\sigma)}\right)^2 \, .
	\label{CondVakh}
\end{equation}
It rewrites as a cubic equation:
\begin{equation}
	\left(g_\mathit{eq}(\sigma) - \frac{2}{3}\right)^3 - \frac{1}{3}\left(g_\mathit{eq}(\sigma) - \frac{2}{3}\right) + \frac{2}{27} - \frac{\sigma^2}{2E\gamma} = 0\, ,
	\label{CondGeqCardan}
\end{equation}
which may have multiple solutions.

\begin{figure}
	\begin{minipage}{0.49\textwidth}
		\centering
		(a)
		
		\includegraphics{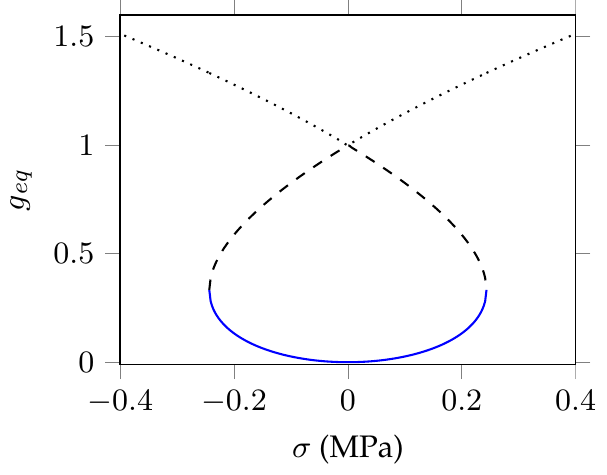}
	\end{minipage}
	\begin{minipage}{0.49\textwidth}
		\centering
		(b)
		
		\includegraphics{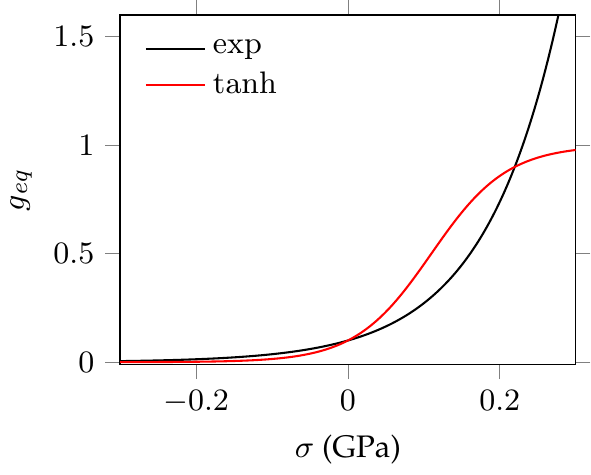}
	\end{minipage}
	
	\caption{Graph of the equilibrium value $g_\mathit{eq}(\sigma)$ in the soft-ratchet model. (a) Roots of (\ref{CondGeqCardan}). The solid line corresponds to a thermodynamically admissible expression of $g_\mathit{eq}$. (b) Classical expressions ``exp'' (\ref{gEqVakh}) and ``tanh'' (\ref{gEqLomb}) of $g_\mathit{eq}$. \label{fig:VakhnenkoGeq}}
\end{figure}

When using Cardano's method, the discriminant
\begin{equation}
	\Delta = \frac{27\,\sigma^2}{4 E^2 \gamma^2} \left(\frac{8E\gamma}{27} - \sigma^2\right)
	\label{CondGeqDelta}
\end{equation}
of the cubic function in (\ref{CondGeqCardan}) is positive if $|\sigma| < \sqrt{8E\gamma/27}$. In this case, the three roots of (\ref{CondGeqCardan}) are real. On figure~\ref{fig:VakhnenkoGeq}-(a), the three real roots are represented, where the parameters are issued from table~\ref{tab:ParamElast}. For comparison, the classical expressions (\ref{gEqVakh}) and (\ref{gEqLomb}) of $g_\mathit{eq}(\sigma)$ are displayed on figure~\ref{fig:VakhnenkoGeq}-(b), where $g_0 = 0.1$ and $\tilde\sigma = 0.1$~GPa. Among the three real roots of (\ref{CondGeqCardan}), only one satisfies $g_\mathit{eq}(0) = 0$ (solid line on figure~\ref{fig:VakhnenkoGeq}-(a)):
\begin{equation}
	g_\mathit{eq}(\sigma) = \frac{4}{3}\sin^2\!\left(\frac{1}{6}\arccos\left(1 - \frac{27 \sigma^2}{4E\gamma}\right)\right) .
	\label{GeqThermo}
\end{equation}
This thermodynamically admissible expression of $g_\mathit{eq}$ is only defined when the discriminant (\ref{CondGeqDelta}) is positive, i.e. for strains smaller than $\sqrt{8\gamma/(27 E)} \approx 2.4\times 10^{-5}$. This bound has the same order of magnitude as (\ref{maxV}).

To summarize, we have shown that the soft-ratchet model is not thermodynamically relevant. A modification of this model has been examined, which results in an implicit definition of $g_\mathit{eq}$ (\ref{CondGeqVakh}). The expression of $g_\mathit{eq}$ is dependent on the choice of a strain energy density function and a storage energy. Furthermore, equation (\ref{CondGeqVakh}) may be hard to solve analytically in some cases. Lastly, the domain of validity of the model may be restricted.

\end{document}